# An Advanced NCRF Linac Concept for a High Energy e⁺e⁻ Linear Collider

K. L. Bane, T. L. Barklow, M. Breidenbach, C. P. Burkhart, E. A. Fauve, A. R. Gold, V. Heloin, Z. Li, E. A. Nanni, M. Nasr, M. Oriunno, E. Paterson, M. E. Peskin, T. O. Raubenheimer and S. G. Tantawi

SLAC National Accelerator Laboratory, 2575 Sand Hill Road, Menlo Park, CA 94025

**Abstract**

We have explored a concept for an advanced Normal-Conducting Radio-Frequency (NCRF) C-band linear accelerator (linac) structure to achieve a high gradient, high power e⁺e⁻ linear collider in the TeV class. This design study represents the first comprehensive investigation for an emerging class of distributed coupling accelerator topology exploring nominal cavity geometries, frequency and temperature of operation. The structure features internal manifolds for distributing RF power separately to each cell, permitting the full structure geometry to be designed for high shunt impedance and low breakdown. Optimized within operational constraints, we find that it is advantageous for the structure to be cooled directly by liquid nitrogen (LN), further increasing the shunt impedance. A crucial part of this design process has been cost optimization, which is largely driven by the cost of peak RF power. The first operation of a distributed coupling structure at cryogenic temperatures and the nominal operating gradient 120 MeV/m is also presented, demonstrating the feasibility of achieving high-gradient performance with a cryogenically-cooled normal-conducting accelerating structure.

## 1. Introduction

TeV scale e⁺e⁻ linear colliders are challenging for many reasons, including their capital and operating costs. We have undertaken an initial study of an alternative accelerator structure with beam characteristics suitable for a TeV collider. This study focusses on a new NCRF structure with internal manifolds distributing the RF to each cell, eliminating the need to transmit RF power through the cavity irises permitting the entire structure to be designed for high shunt impedance and low breakdown. In addition, the structure will be cooled to ~77 K, increasing the shunt impedance by ~2.5 and reducing the breakdown rate.[1] The optimal gradient and length for a NCRF linac depends directly on the cost of RF power, usually characterized by cost per peak RF kilowatt, which includes both the modulator and the RF source. DOE-HEP funds a General

Accelerator R&D (GARD) program, which has produced a decadal roadmap that includes a cost goal of $2/peak KW.[2] We *assume* that cost in the linac concept presented here. This study is limited to the accelerator and does not include the necessary sources, damping rings, or beam delivery systems.

To guide the design, we scale the accelerator requirements for luminosity and beam power from the existing established CLIC[3] and ILC[4] designs; see Figure 1. Table 1 shows the main parameters for a 2 TeV center of mass linac. For lower energy designs, parameters in Table 1 marked with a [†] should scale roughly linearly, with modifications below 300 GeV to maintain a linear luminosity scaling with energy. The physics case for a TeV scale $e^+e^-$ collider is described in Appendix A.

## 2. Cost Optimization

The primary obstacle to building any next-generation $e^+e^-$ collider appears to be the cost. The approach of this proposal is therefore to seek the minimum for both capital and operating costs. For an NCRF machine, the dominant capital cost is the RF sources. In practice, an NCRF collider operates with a low duty cycle where the cost of RF sources is driven by the peak power that the RF system must deliver to the linac. RF sources optimized for short pulse and low duty cycle operation can have significantly simplified cooling systems. Other potential simplifications in the design can, for example, reduce susceptibility to oscillations in the RF source. The GARD RF decadal roadmap specifically lays out these requirements. In particular, it calls for a dramatic improvement in the cost of RF sources (in $/peak kW) for low duty cycle. The GARD RF power goal is a cost of $2/peak kW encompassing the full system including the modulator and high-power amplifier. As detailed in the report, this cost represents an order of magnitude decrease from the current commercial state of the art for the frequencies (C/X-band) and efficiencies explored in this design study. We take this cost as an assumption for the main linac design and solve for the major parameters given in Table 2 through the overall system optimization. The impact of the pulse format on the capital and operating costs is also considered. The repetition rate is kept low at 120 Hz to simplify the cooling of detector electronics and to simplify the damping ring. For the beam format, we have evaluated both individual RF pulses and RF pulse trains repeated at the repetition rate. We found that individual pulses provided reasonable pulse lengths at ~0.25 microseconds. Beam loading has also been optimized to reduce the operational cost of the accelerator, trading increased peak power requirements for higher electrical efficiency and reduced cooling capacity. The temperature of operation (at both 77 K and 300 K), frequency of operation (C/X-band) and iris aperture were also optimized. A description of the optimization procedure for the accelerator is described in detail in the Appendix B. Further comparisons of the RF performance of C/X-band structures is provided in Appendix C.

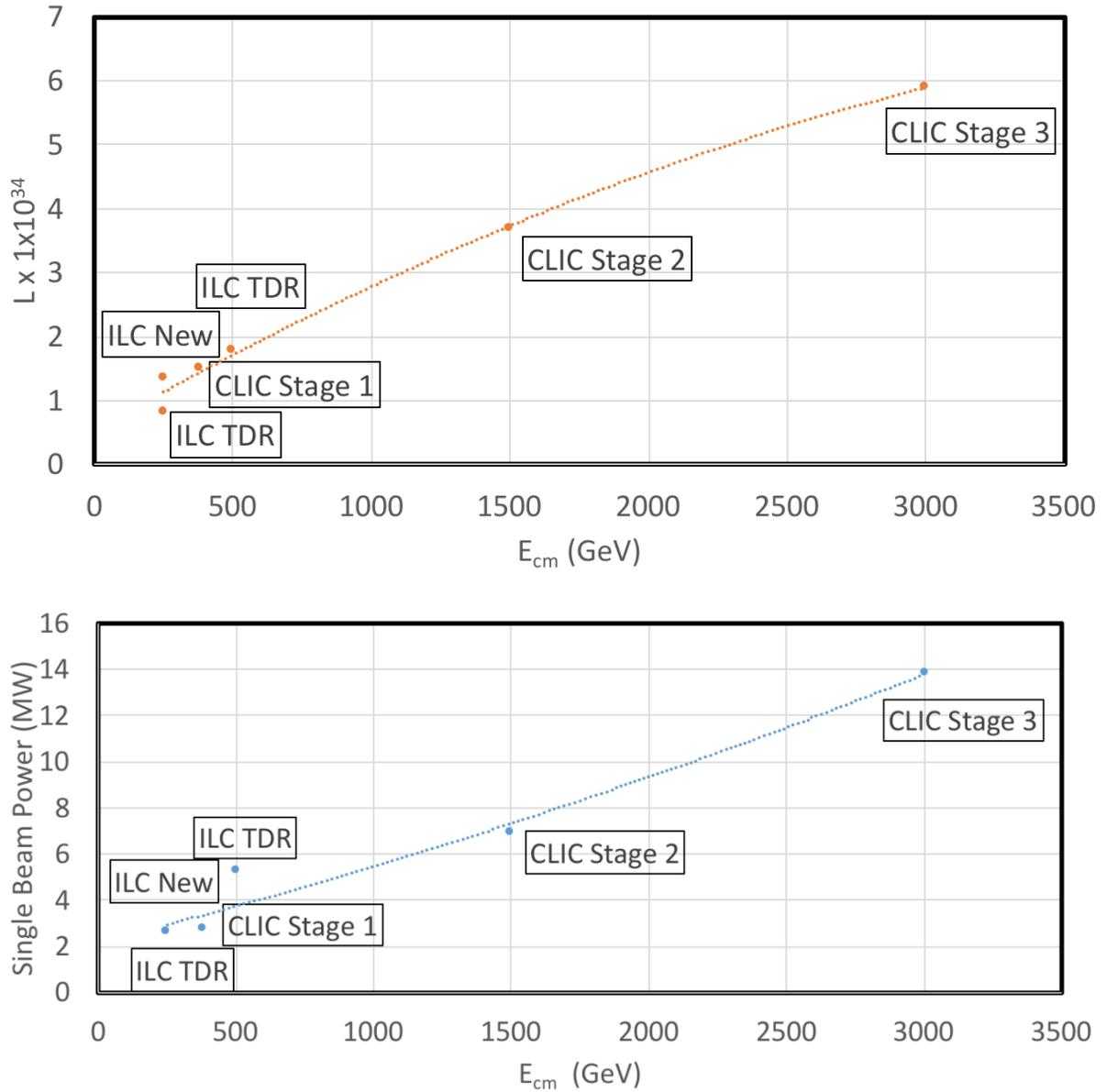

**Figure 1:** (above) Linear collider luminosity (*L*) *circa* 2017 and (below) single beam power required as a function of Center of mass energy.[5]

| Parameter | Symbol | Units | Value | Notes |
|---|---|---|---|---|
| Center of Mass Energy | | TeV | 2 | †Denote parameters that scale with center of mass energy |
| Single Beam Energy† | $E_b$ | GeV | 1000 | |
| Single Beam Power† | $P_b$ | MW | 9 | |
| Train Rep Rate | | Hz | 120 | See Figure 2 |
| Bunch Charge | N | x10$^{10}$ | 0.625 | 1 nC |
| RF Pulse Length | | ns | 250 | |
| Bunch Spacing | | periods | 19 | 3.3 ns |
| Average Current | | μA | 9 | |
| Peak Current | | A | 0.3 | |
| Luminosity† | L | x10$^{34}$ cm$^{-2}$s$^{-1}$ | 4.5 | |
| Operating Temp | $T_{op}$ | K | 77 | |
| Linac Gradient | | MeV/m | 117 | |
| Filling Factor | | | 0.9 | |
| Single Linac Length† | | km | 9.5 | |
| Cavity Fundamental | | GHz | 5.712 | |
| Shunt Impedance | | MΩ/m | 298 | |
| a/λ | | | 0.05 | |
| Beam Loading | | % | 42 | |

**Table 1:** Parameters for the main linac

To make the optimization practical, we have assumed conservative values of the outfitted tunnel cost informed by the ILC TDR, and conservative values for the structure cost based on recent SLAC experience. For this point design, we have selected a C-band structure at an operating temperature of 77 K that reduced the overall capital cost of the collider. The pulse format is shown in Figure 2. Single 250 ns RF pulses are repeated at 120 Hz, with 75 x 1 nC electron bunches per pulse at a spacing of 19 periods. The large number of periods separating each bunch aids in suppression of long-range wakefields. In this estimate, a 1 TeV linac with 9 MW beam power would cost 3.2 G$. For reference, the cost of the linacs in the ILC TDR[6] design (2 x 250 GeV) is estimated at 5200 2012 MILCU. Therefore this is, very roughly, a savings of a factor of 3 per GeV.

| Parameter | Units | Value |
|---|---|---|
| RF Source Cost | $/peak kW | 2 |
| Temperature | K | 77 |
| Main Linac Cost | G$/TeV | 3 |
| Structure Cost | k$/m | 100 |
| RF Compressors | k$/m | 15 |
| Beam Loading | % | 42.5 |
| Gradient | MeV/m | 117 |
| Pulse Length | ns | 250 |
| Cryogenic Load at 77 K (See Sec. 5) | MW | 24.4 |
| Est. AC Power (two linacs - see Table 4 for subtotals) | MW | 342 |
| Est. Electrical Power for Cryogenic Cooling | MW | 170 |
| RF Source efficiency (AC line to linac) | % | 50 |

| Train Repetition Rate | Hz | 120 |
|---|---|---|
| Additional Linac Length (Instrumentation, Cryogenic Feeds, etc.) | % | 10 |
| Tunnel Cost | k$/m | 50 |

**Table 2:** Parameters for 2 x 1 TeV accelerator (excludes final focus, IP, injector, damping ring and pulse compressors).

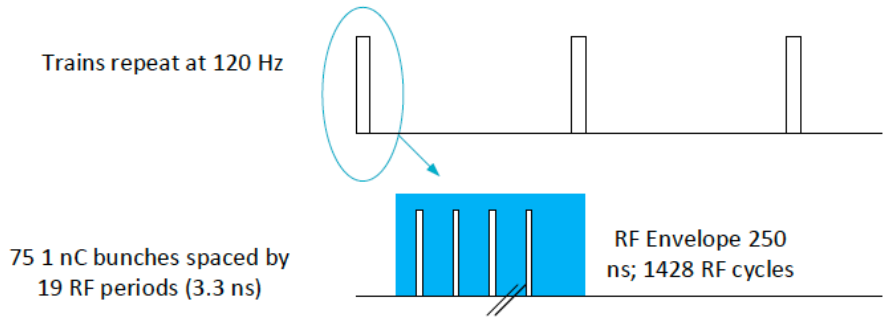

**Figure 2:** Beam format for the main linac showing the 250 ns flat top of the RF envelope.

### 3. Cavity Structure

Linacs accelerate charged particles using oscillating electric fields excited within RF cavities (cells) that are joined together to form a beamline. Typically, RF power is fed to the linac from one point and flows through adjacent cells using coupling holes that also serve as beam tunnels for charged particles.[7] This approach is similar to the designs for both the NLC and CLIC. Consequently, the linac design process requires careful consideration of the coupling between adjacent cells.[8] This limits the ability of designers to optimize the cell shape for efficiency and or gradient handling capability.

A distributed coupling accelerator topology[9,10] is selected for this design in order to maximize the accelerator efficiency and to achieve high gradient. In distributed coupling accelerators, the fundamental accelerating mode of each RF cavity is individually powered and effectively isolated from adjacent cells by a small beam aperture. Each cell can be optimized under fixed constraints, such as peak electric field to accelerating gradient, to maximize the shunt impedance.[11] In addition, the phase advance per cell can also be treated as a free variable.

Preliminary studies on single-bunch short range wakefields indicate that for the nominal charge of 1 nC and an operating gradient at or above 100 MeV/m, an iris aperture radius as small as 2.62 mm is tolerable (see Appendix C). Harmonic frequencies of 2.856 GHz were investigated for cavity designs. C-band and X-band structures achieved similar shunt impedances if the period of the cell was unconstrained. The C-band structure, shown in Figure 3, was selected due to the slightly lower cost of RF power for lower frequencies. Although not the cell period producing the absolute maximum achievable shunt impedance, a cell period of λ/3 was selected to simplify the RF distribution manifold required to power the structure. The room temperature and cryogenic shunt impedances of the structure are 133 MΩ/m and 298 MΩ/m, respectively. A summary of the structure parameters is given in Table 3. This cavity structure is the baseline for the proposed accelerator and requires further detailed design for coupling and RF distribution, as discussed below. Additionally, a modified structure that includes both damping and detuning will be required to suppress long-range wakefields, as shown in Section 4.

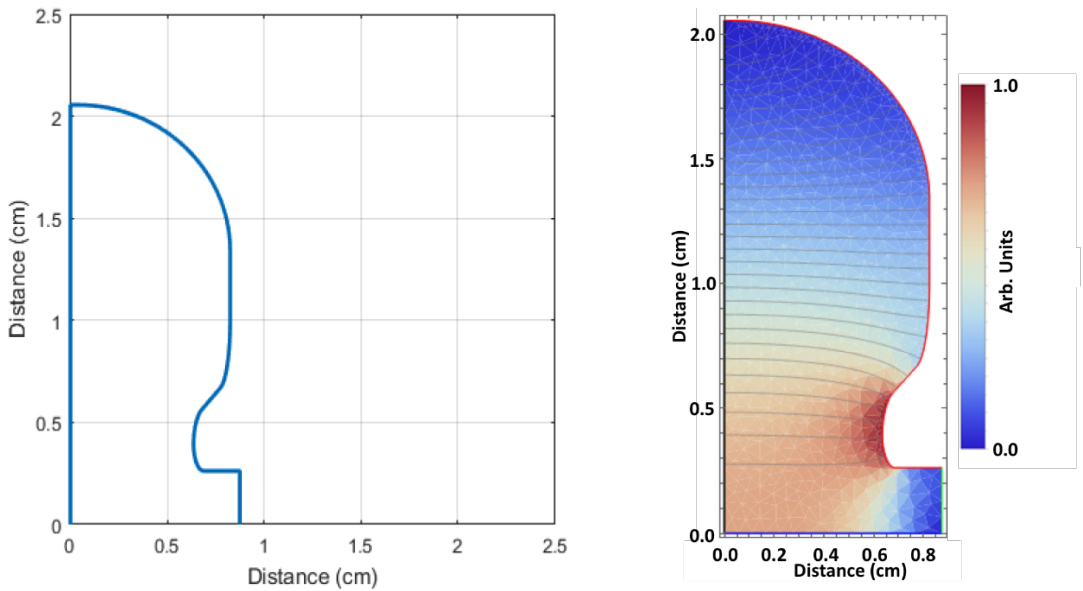

**Figure 3:** (left) Geometry and (right) electric fields in a C-band cavity of length λ/3.

At nearly 300 MΩ/m for its planned mode of operation, this structure stands in stark contrast to the structures evaluated for NLC and CLIC, even with X-band operation. NLC structures operated with a shunt impedance of up to 98 MΩ/m,[12] albeit with a larger aperture (3.75 mm). CLIC structures currently under consideration operate with a shunt impedance of 95 MΩ/m with an aperture of (3.15-2.35 mm).[13,14] The prospect of a significantly higher shunt impedance dramatically reduces the cost of operation (from the reduced peak RF power requirements), increases the nominal operating gradient and significantly reduces the peak current required to operate with heavy beam loading. For this point design, the beam loading or fraction of steady state RF power absorbed by the electron beam is 42%. Reducing the peak current also allows a longer bunch spacing of 3.3 ns for 1 nC bunches.

The distributed coupling topology provides additional benefits when operated with heavy beam loading. In particular, the RF distribution manifold powering the cells must be overcoupled to operate efficiently, which reduces the fill time of the structures. A fill time for the structure equal to or less than the flat top of the gradient or the electron bunch train duration (250 ns) is desirable, as it will help to minimize the additional thermal load. This is possible due to the heavy beam loading of the accelerator, which affords us the ability to design the coupling to the cavities such that they are highly overcoupled when no beam is present during the filling and decaying portion of the RF pulse.

The high shunt impedance and heavy beam loading also reduce the RF pulse compression needed to achieve the peak power during the cavity fill time. Options for the RF pulse format and the optimal design are discussed in Appendix D. Operating at the nominal gradient of 117 MeV/m with 42% beam loading requires only 80 MW of RF power delivered directly from the RF source per meter of structure.

| Parameter | Units | Value | Notes |
|---|---|---|---|
| Train Rep Rate | Hz | 120 | See Figure 2 |
| Bunch Charge | x10$^{10}$ | 0.625 | 1 nC |
| Bunch Length | μm | 100 | |
| Bunch Train Duration | ns | 250 | |
| Bunch Spacing | periods | 19 | 3.3 ns |
| Operating Temp | K | 77 | |
| Linac Gradient | MeV/m | 117 | |
| Cavity Fundamental | GHz | 5.712 | λ=5.25 cm |

| Shunt Impedance | MΩ/m | 298 | |
|---|---|---|---|
| a/$\lambda$ | | 0.05 | a=2.62 mm |
| $E_{max}/E_a$ | 2.12 | | |
| Beam Loading | Percent | 42.5 | |

**Table 3:** Structure Parameters

Any concept for a distributed feeding network is required to simultaneously provide power and the appropriate phase advance for each cell. In this case, the cells are identical and require equal power and a constant phase advance. For a typical electron linac with a particle moving at nearly the speed of light, the phase advance per cell is $2\pi P/\lambda$; where $P$ is the periodic separation between cells and $\lambda$ is the free space wavelength.

With the exception of a dielectric-free coaxial line, this phase advance cannot be provided by any form of wave guiding structure, which always have a guided wavelength $\lambda_g > \lambda$. Alternatively, one can use waveguides oriented so that the center $E$-plane is the same as the $E$-plane of the accelerator cell.[21] Because the guided wavelength does not match the free-space wavelength, one can imagine solutions where the distribution waveguide is bent like a serpent to achieve the appropriate phase advance. This is valid, but one can use more than a single manifold. A natural interval in spacing for tapping into the RF waveguide manifold with power couplers, as shown below, is every $m\lambda_g/2$. Here, $\lambda_g$ is the guided wavelength within the manifold and m is an integer. Therefore, for a structure with a phase advance/cavity of $2\pi/n$, one can use n manifolds where each one of them is being tapped every $\lambda_g$. Hence the phase advance is also an optimization parameter.

A $\pi$ phase advance per cavity is a special case that simplifies the design of the system. In this case, n=2 and only two manifolds are needed. Initial distributed coupling structures shown in Figure 4 have this mode of operation, which is close to being optimal. For the large-scale optimization of a collider structure with small beam apertures, a phase advance of $2\pi/3$ is slightly better in terms of shunt impedance. An RF distribution manifold will therefore require some serpentine waveguides, Figure 5. As described, each segment of the distributed-coupling accelerator structure can be manufactured from two blocks as shown in Figure 4. The structure is designed so there are no currents crossing the mid-plane along the long dimension. This reduces the complexity of manufacturing the structure and provides logical places for both the cooling manifolds and the tuning holes.

The manifold consists of a set of cascaded T-junctions with power going through the tap-off ports. The cavity coupling port is designed to give a matched port when the cavity is loaded with the design current.

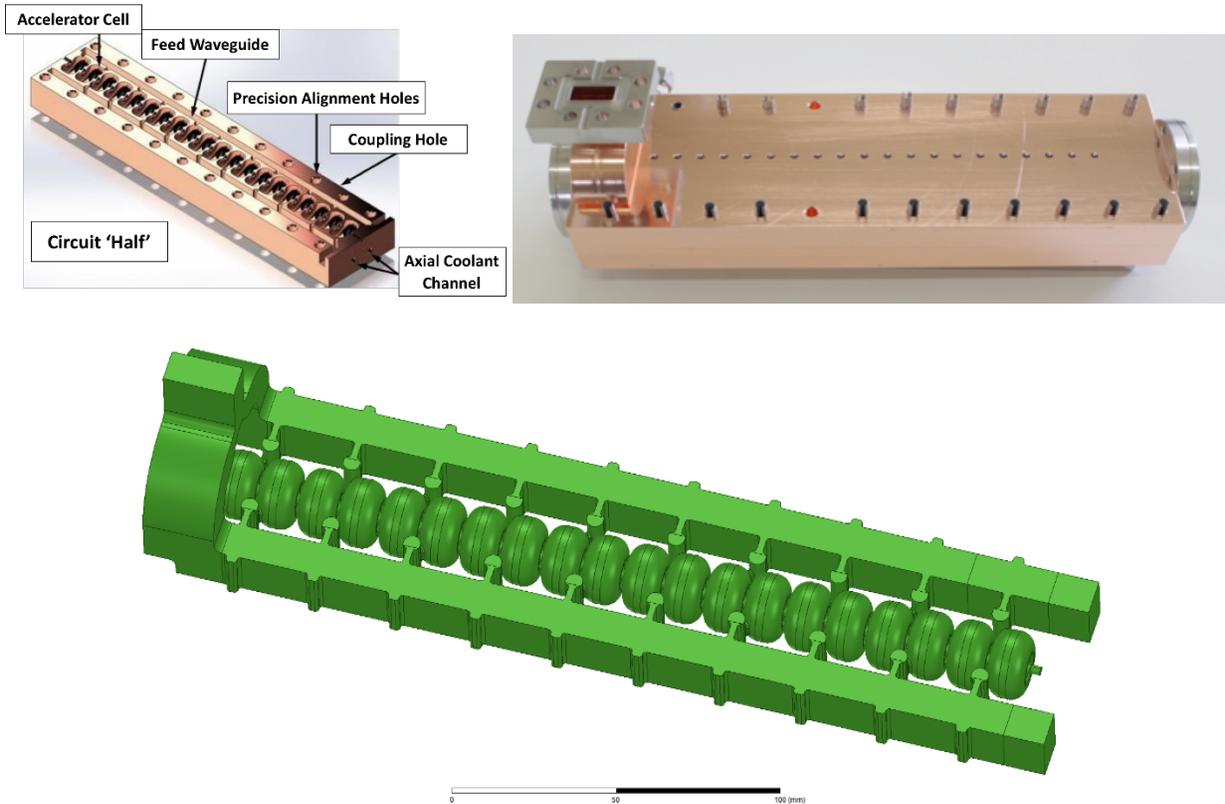

**Figure 4:** (Top-left) Illustration and (top-right) photograph of the distributed coupling linac constructed from two quasi-identical copper blocks, including the tuning pins, alignment holes and cooling channels. (Bottom) Solid model of the vacuum space for the RF manifold and cavities of an X-band $\pi$ phase advance structure.

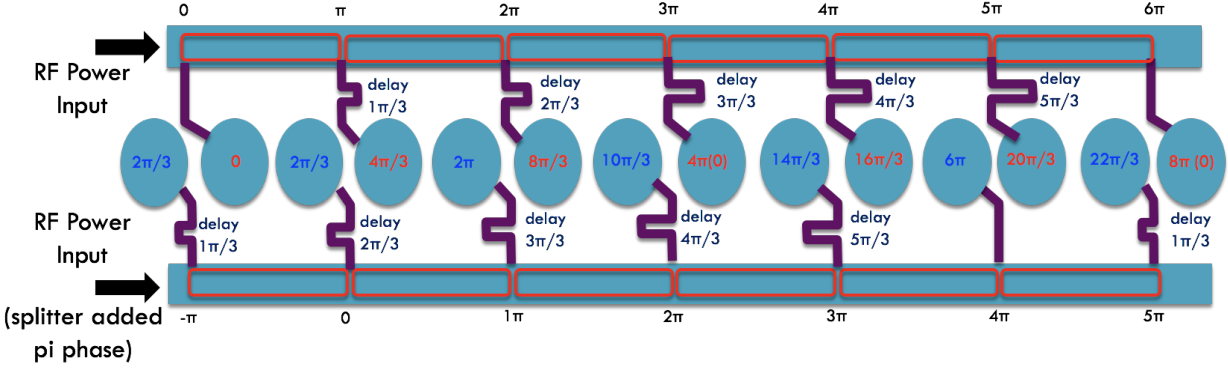

**Figure 5:** Schematic of the serpentine RF distribution for the proposed parallel feed waveguide $2\pi/3$ C-band structure.

## 4. Wakefields

Wakefields and the resulting emittance degradation are of significant concern for any practical collider design. A complete wakefield analysis requires advanced beam dynamics simulations that account for both short-range and long-range effects. They must include operational machine tolerances, the design and optimization of wakefield damping, and cell detuning to mitigate long-range wakes. While a complete analysis and design is beyond the scope of this work, an initial effort to ascertain the validity of this structure was undertaken using the performance characteristics of existing collider designs.

### a. Short-Range Wakefield

We began with the short-range wakefield for the structure. The small iris aperture of $a/\lambda=0.05$ has the potential for significant short range wake effects. The emittance growth in the structure was analyzed for various beam parameters by varying the in offset phase for longitudinal wake suppression and the bunch length for transverse wake suppression. The goal was to achieve a residual energy spread in the range of $\sigma_{\delta 0}$ = 2-5x10$^{-3}$ and a maximum of 1% correlated energy spread for Balakin-Novokhatski-Smirnov (BNS) damping.

The short-range wakefield study analyzed both the actual cavity profile and an approximate profile, shown in Figure 6(left), for numerical and analytical calculations, respectively. It was found that the actual and approximate cavity profiles for both the longitudinal loss factor (Figure 6 middle) and transverse loss factor (Figure 7 left) were in close agreement. The resulting longitudinal and transverse wakefields with a $\sigma_z$ = 150 $\mu$m are shown in Figure 6 (right) and Figure 7 (right), respectively.

To assess the effectiveness of BNS damping in mitigating the transverse wakefield,[15] we make several assumptions. For each linac, we assume an injection energy of 10 GeV with a $\beta_{x0}$ = 4 m, and a final energy of 1 TeV with an increase in $\beta \sim E^{\frac{1}{2}}$. Simulations are performed assuming smooth focusing with a 1 nC Gaussian bunch with 51 slices representing $\pm 2.5\sigma_z$. The bunch is offset in normalized coordinates, with an initial offset of $x_0 = \sigma_x$. The centroids of all slices for the resulting phase space in x and $\beta$x' are plotted in Figure 8 with dots for two values of $\sigma_z$: 150 $\mu$m (left) and 100 $\mu$m (right). The circle shown in Figure 8 gives the amplitude of the initial offset. The correlated rms energy spread is 1% (tail lower energy than head). The head of the bunch can be identified residing on the circle defined by the initial offset as it does not experience any wakefield. These plots give emittance amplification factor of 4.5 (left case) and 0.76 (right case). We observe that a $\sigma_z$ of 100 $\mu$m results in small emittance growth due to an initial transverse offset.

The initial beta-function of $\beta_{x0}$ = 4 m could be provided with a 4-m FODO cell having 90-degrees phase advance.  At 10 GeV, this would require 12 cm quadrupoles with a gradient of 200 T/m.  To maintain the scaling of the $\beta$-function with $E^{\frac{1}{2}}$, the cell and quadrupole lengths would scale in the same manner and the phase advance would remain 90-degrees.[16]  The high quadrupole gradient may be most easily achieved with variable strength permanent magnet quadrupoles; a recently constructed device demonstrated 201 T/m with a 10 mm bore and 45% variability.[17] A possible permanent magnet material that performs well at 77 K[18,19] is $Pr_2Fe_{14}B$, known as VD131TP and VD131DTP and made by Vacuumschmelze.[20]  An advantage of permanent magnets is the lack any water-induced vibrations from cooling or local boiling in the liquid Nitrogen.

To mitigate the longitudinal wakefield, the bunch needs to run off-crest with an average offset in phase. The final offset in phase needs to be determined in conjunction with the correlated energy spread. A calculated residual energy spread of $\sigma_{\delta 0}$ = $2\times10^{-3}/4.8\times10^{-3}$ is achieved with a 21/0 degree average phase offset (ahead of crest) and a modest/negligible 6.5%/0% reduction in gradient.

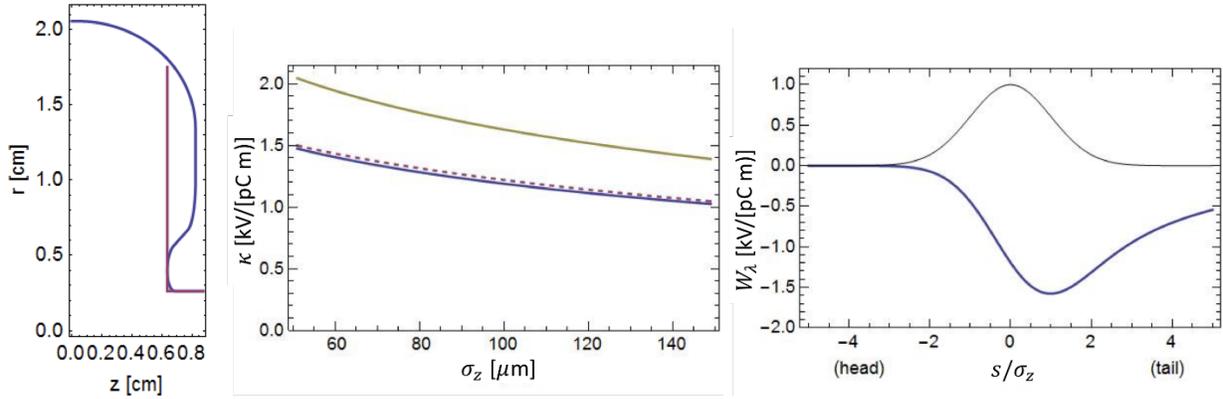

**Figure 6:** (left) Actual (blue) and approximated (red) structure profile. For the approximated structure a period $p$ = 1.75 cm, a gap $g$ = 1.272 cm, and an iris radius $a$ = 0.262 cm were used. (middle) Longitudinal loss factor (blue), analytical (dashed), alternate $\pi$-mode (C-band, $a/\lambda$ = 0.05) structure (yellow) vs rms bunch length. (right) Longitudinal bunch wake (blue) and bunch shape (black) with head to left. Result shown for an rms length $\sigma_z$ = 150 μm.

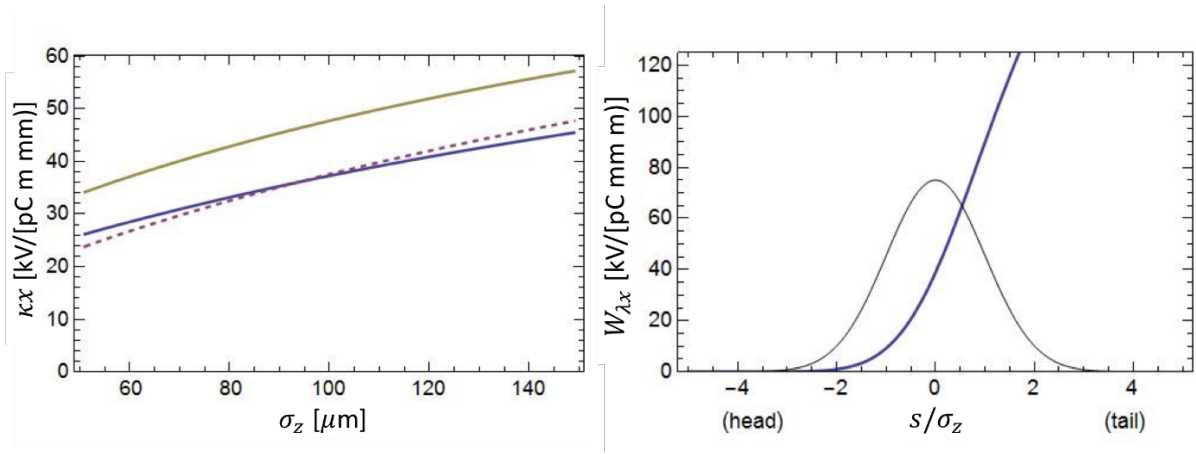

**Figure 7:** (left) Kick factor (blue), analytical (dashed), alternate $\pi$-mode (C-band, $a/\lambda$ = 0.05) structure (yellow) vs rms bunch length. (right) Transverse bunch wake (blue) and bunch shape (black) with head to left. Result shown for an rms length rms length $\sigma_z$ = 150 μm.

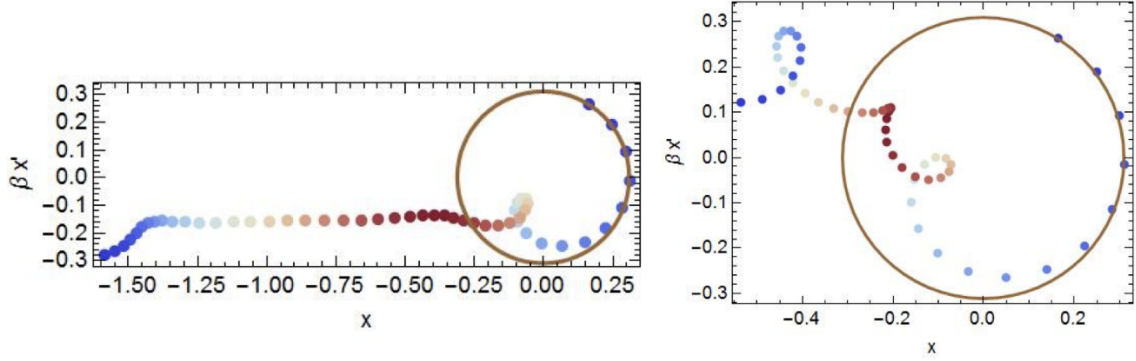

**Figure 8:** For C-band, $a/\lambda$ = 0.05, $2\pi/3$ cavity, 1% BNS correlated energy spread, $\beta_{x0}$ = 4 m: $\sigma_z$ = 150 μm (left) and $\sigma_z$ = 100 μm (right). The projected emittance growth factors are 4.5 (left) and 0.76 (right). Slices are color-coded with size of $\lambda_{zi}$, where large (small) is red (blue).

### b. Long-Range Wakefield

Damping of the long-range wakefield was the subject of extensive investigations for collider designs such as NLC and CLIC. Both of these accelerator designs have incorporated cell-to-cell detuning and damping for wakefield suppression.[21] Detuning works by preventing the coherent addition of dipole modes along the length of the structure. In order to achieve detuning, small modifications must be designed and incorporated into each cell of the structure. The modifications require simulation and design with high-performance computing tools that have been developed for linear collider structure design.[22] The efficacy of detuning is limited by the eventual recoherence of the wakefield modes in the structure. Therefore, broadband damping must be added to suppress the wakefields. In order to symmetrically suppress both orientations of the dipole mode, structures have been designed with four apertures in each cell that couple either to individually damped waveguides[23] containing an absorbing material (SiC) or to common manifolds that transport the power away from the structure to a common load.[24]

Understandably, in the actual structure for the main linac presented in this design, these two well-known means of wakefield suppression will need to be incorporated in two ways. First, by detuning the dipole frequency in a Gaussian density distribution, the wakefield can be dramatically suppressed in a short distance, e.g. at the second bunch of the bunch train. Second, with damping to suppress the wakefield at a longer distance from recoherence of the detuned modes, ideally through the existing manifold and an additional one for the second orientation of the dipole mode.

The most immediate concern for long-range wakefields are the dipole modes. The dominant contribution to the long-range dipole wakefield is from the first 2 to 3 dipole passbands of the structure. The contribution of each mode is quantified by the kick factor determined by the structure's geometry. As a function of distance, the long-range dipole wakefield is calculated as the sum of the sinusoidal spatial oscillations with wavenumbers $\omega n/c$, scaled by the kick factors.[25] The dipole wakefield was calculated for a 20-cell C-band structure. The 20 cells are identical and there is no de-tuning incorporated (see Figure 9). While the wakefield of such a structure does not represent the final design for a linear collider, the calculation presented here provides a realistic peak value for comparison with other designs.

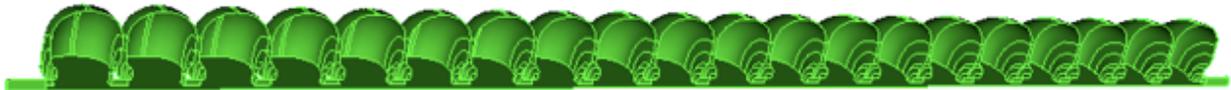

**Figure 9:** A 20-cell structure with uniform cells utilized for long-range wakefield calculations.

Figure 10 shows the dipole wakefield for a uniform 20-cell C-band structure. Each cell is 1/3 of a wavelength and the total structure length is 351 mm. A $Q_{damp}$ of 1000 was used for all the dipole modes. The peak value of the transverse wakefield at the origin is about 50 V/pC/mm/m. This value is about half that of the X-band structure proposed for the NLC. The bunch spacing in consideration is 19 periods of the C-band wavelength, which is about 1 m. Based on the X-band structure studies for the NLC, about 10% detuning of the dipole modes would be needed to minimize the wakefield at the first subsequent bunch by a factor of >50. The exact value will depend on the number of modes used for the detuning.

In Figure 11 we see that for the first dipole band, a four sigma Gaussian detuning does provide suppression for the first subsequent bunch located at s = 1 m. With the decoherence optimized for one meter, a recoherence of the wakefield is observed by ~75 m. This distance is close to the length of the bunch train. Simulations shown in Figure 12 introduce a damping mechanism that reduces the quality factor for the dipole modes to 1000, completely suppressing the recoherence. This damping mechanism is not yet designed. One possible solution is utilizing the parallel feeding waveguide, Figure 4 and Figure 5, for the C-band design which can naturally provide manifold damping for the dipole modes. Additional damping schemes can be added to achieve the needed damping for both x-y polarizations.

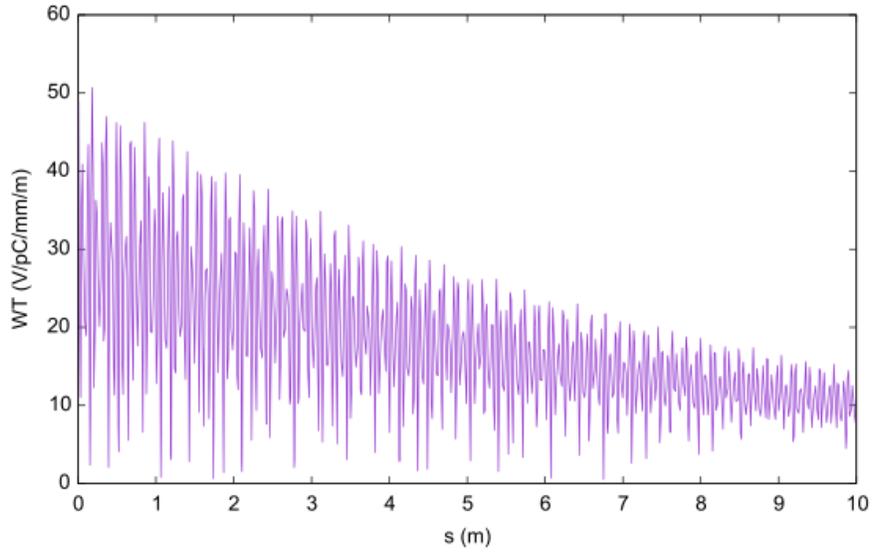

**Figure 10:** Long-range transverse dipole wakefield envelope for the uniform 20-cell C-band structure shown in Figure 9.

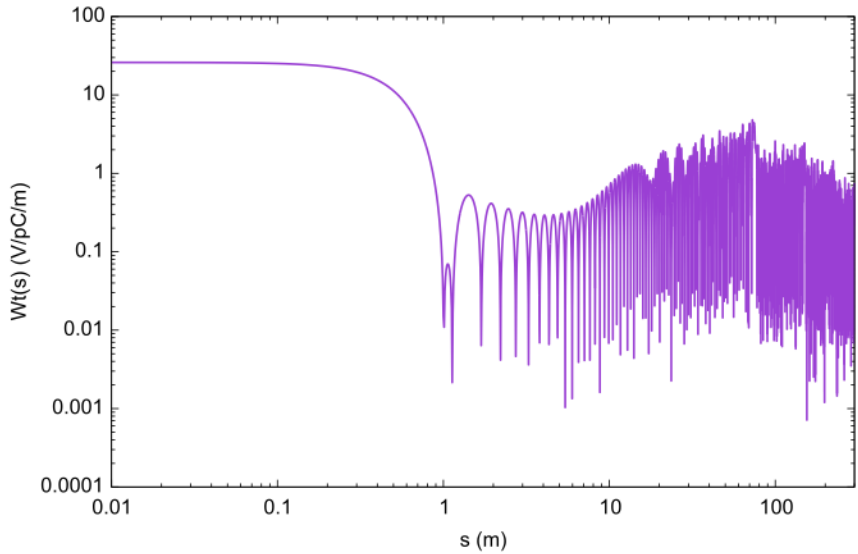

**Figure 11:** Long-range wakefield calculation for the first dipole band with cell detuning incorporated. In this simulation a four-sigma detuning is utilized over 80 cells of structure (~2 m), centered at a frequency of 9.5 GHz with a $\Delta f / f_c$ of 5.6%. These simulations include ohmic losses from the copper walls of the structure.

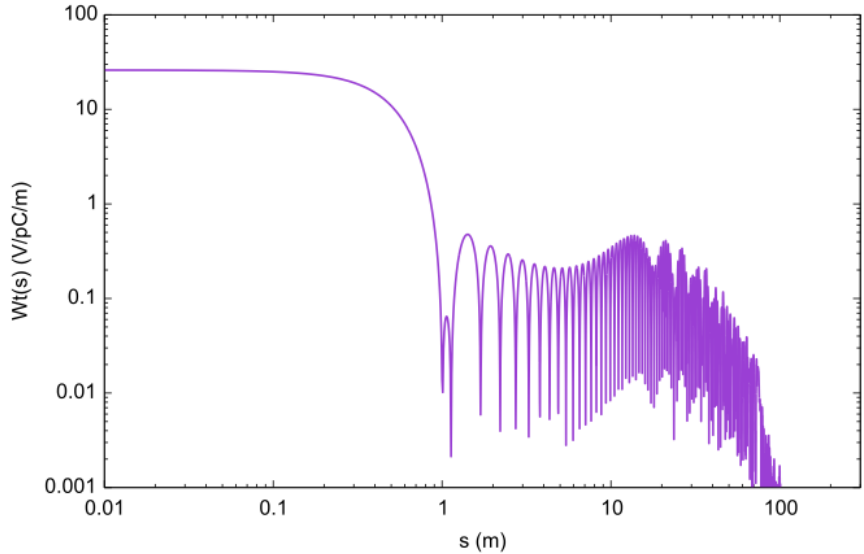

**Figure 12:** Long-range wakefield calculation for the first dipole band with cell detuning and an artificial damping factor incorporated. The damping reduces the qualify factor of the dipole modes to 1000. The detuning parameters are the same as the simulations in Figure 11.

## 5. Cryogenic Cooling Complex

The cooling load at cryogenic temperature is 12.2 MW for each 1.0 TeV linac, with the operating parameters assumed in the previous section. The linac has a length of ~9400 m including a 10% allowance for instrumentation sections, cryogen feeds, *etc.* The cooling utilizes the latent heat of vaporization of LN of 199 kJ/kg. The thermal power deposited in the accelerating structures is derived from the difference between the RF power and the electron beam power, with the parameters given in Table 4. Thus, the required cooling is ~0.5 W/cm$^2$, which is safely in the nucleate boiling regime for liquid nitrogen at 1 bar. The expected $\Delta$T is ~2 K. The critical heat flux for the transition to film boiling is above 10 W/cm$^2$.[26] The vibrations generated by the nucleate boiling must also be characterized. The electrical power requirements for the RF and cryogenic cooling system are also listed in Table 4 and are based on a 15% overall conversion efficiency for the liquid nitrogen cryoplant.[27]

The simplified cross section of the cryomodule is shown in Figure 13. The cryomodule design is rather conceptual and has not been optimized. The cryostat consists of two coaxial tubes with a separate insulating vacuum between them. The accelerator structure and quadrupoles are completely submerged in the liquid nitrogen. For a nitrogen vessel with a 30 cm radius, and a liquid depth of 20 cm, the liquid flow rate at the delivery junction is 3.7 L/s, and has a flow velocity of 0.05 m/s. The pressure of the gas is ~1 bar. The gas flow velocity has a maximum value of 5.4 m/s; and the maximum pressure drop of 5 mbar. The waveguide feed into the cryostat is shown in Figure 13 (right). There is one feed per accelerator section. The upper section of waveguide will be copper plated stainless steel, with a heat leak of ~5 W. It has two welded flanges to permit assembly after the rafts are inserted in the cryostat. A longitudinal section of the full cryostat is shown in Figure 14. It is 8.9 m long, and is intended to couple directly to adjacent cryomodules with welded sleeves. We expect the tolerances described in the CLIC CDR Table 2.10[28] for alignment and vibration will be applicable to this design as the structure aperture and bunch charge are similar.

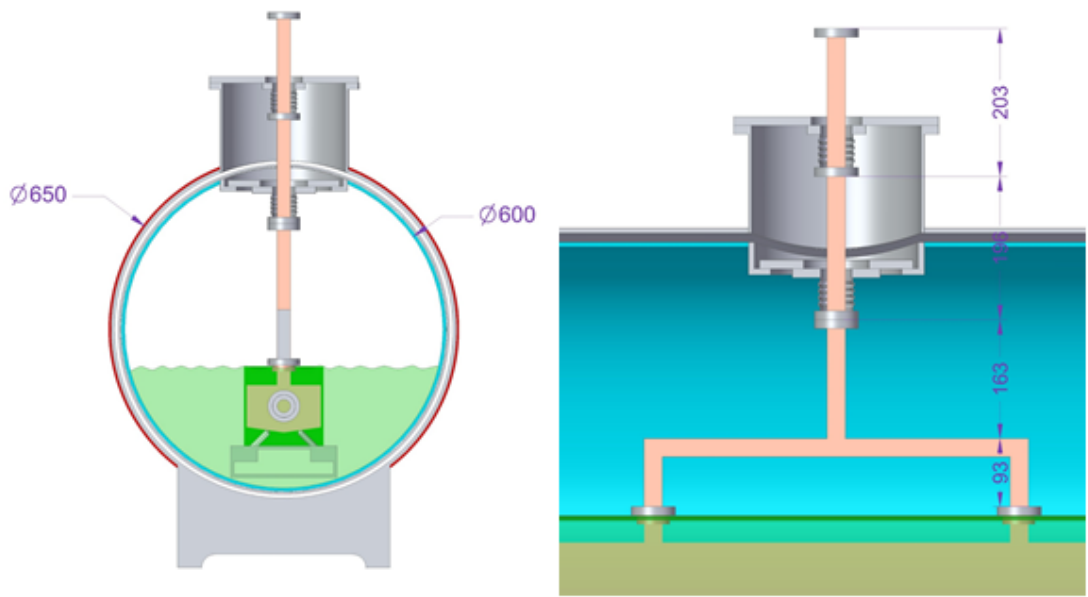

**Figure 13:** A cross section of the accelerator cryostat (left) and a detailed view of the waveguide into cryostat (right).

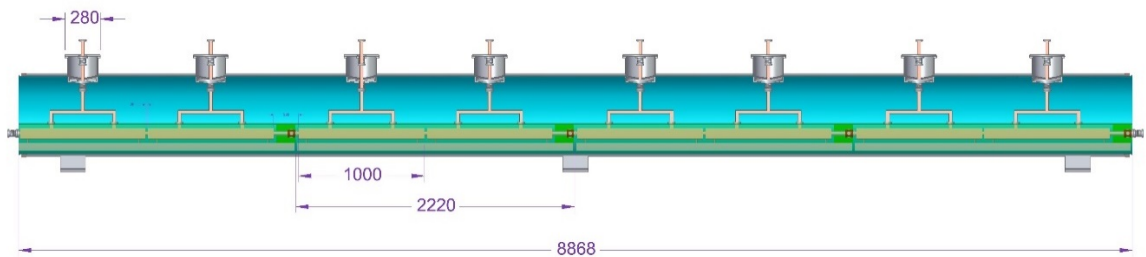

**Figure 14:** A view of the longitudinal cross section of the full cryomodule.

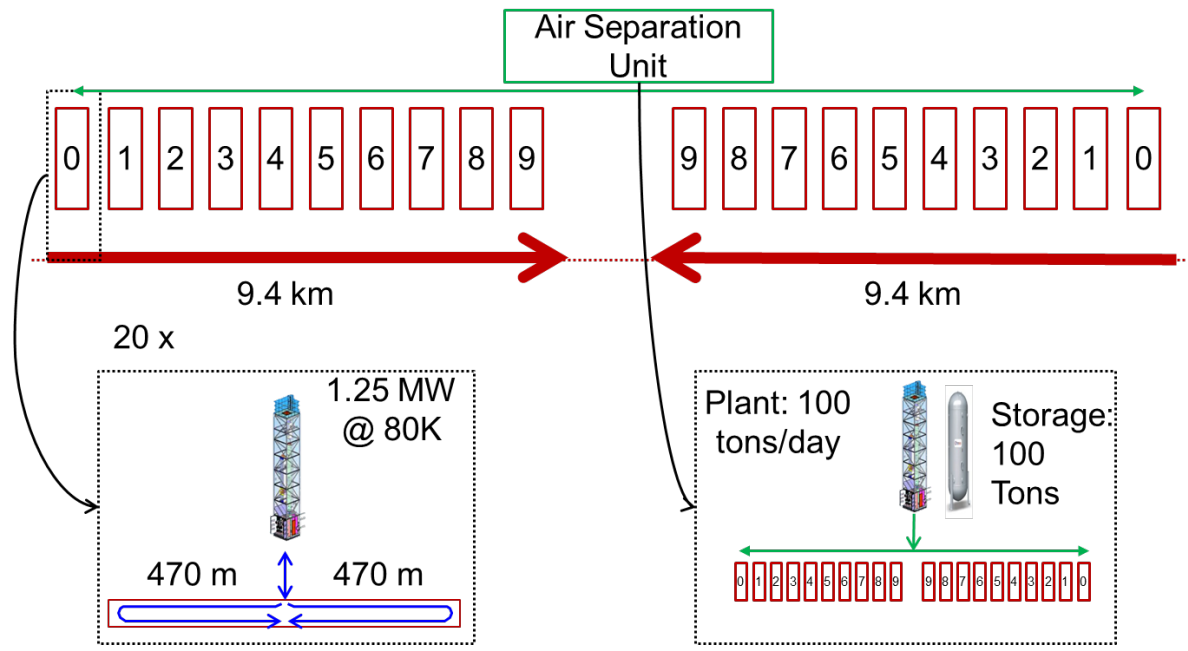

**Figure 15:** Simplified schematic of the overall cryogenic system. The cryogenic system consists of 20x 1.25 MW 90 K refrigeration plants requiring ~7 MW of electrical power each, and one air separation requiring ~2 MW capable of producing 100 tonnes of LN/day.

The layout of the reliquification plants is shown in Figure 15. Each plant has a refrigeration power of 1.25 MW, so each linac requires 10 plants. The LN runs in both directions along the linac, with the length of each run being ~470 m. The power per run is 0.6 MW and the mass flow of LN into each run is ~3 kg/s. In addition, an Air Separation Unit (ASU) with a capacity of 100 tonnes/day provides startup and makeup liquid. The linac LN inventory is ~1600 m³.

| Parameter (2 TeV CoM) | Units | Value |
|---|---|---|
| Nitrogen Reliquification Plant Cost for Cryogenic Cooling | M$/MW | 18 |
| Single Beam Power (1 TeV linac) | MW | 9 |
| Total Beam Power | MW | 18 |
| Total RF Power | MW | 86 |
| Heat Load at Cryogenic Temperature | MW | 25 |

| Electrical Power for RF | MW | 172 |
|---|---|---|
| Electrical Power for Cryo-Cooler | MW | 170 |

**Table 4:** Parameters and power load required for the cryogenic cooling systems.

## 6. High-Gradient Test of Cryogenically-Cooled Accelerating Structure

X-band distributed coupling accelerator structures using multiple fabrication methods have been built and room temperature tested at SLAC[29] with high power RF and with beam[30] at the X-band Test Area in NLCTA.[31] These experiments have demonstrated the structures' abilities to achieve gradients beyond the requirements envisioned here while meeting the designed RF performance. An X-band structure made of hard copper fabricated with e-beam welding has been installed in a LN cryostat and tested with low power RF. The structure's RF performance is shown in Figure 16 at both room temperature and 77 K. The intrinsic $Q_o$ of the structure measured independently from the S-parameters at both temperatures increased from ~11,000 to 27,500 or an increase of 2.5, which exceeds the assumed increase used for the design presented here. Future measurements will explore the role of material stock and thermal cycling in fabrication (e.g. from brazing) in achieving the largest increase in shunt impedance.

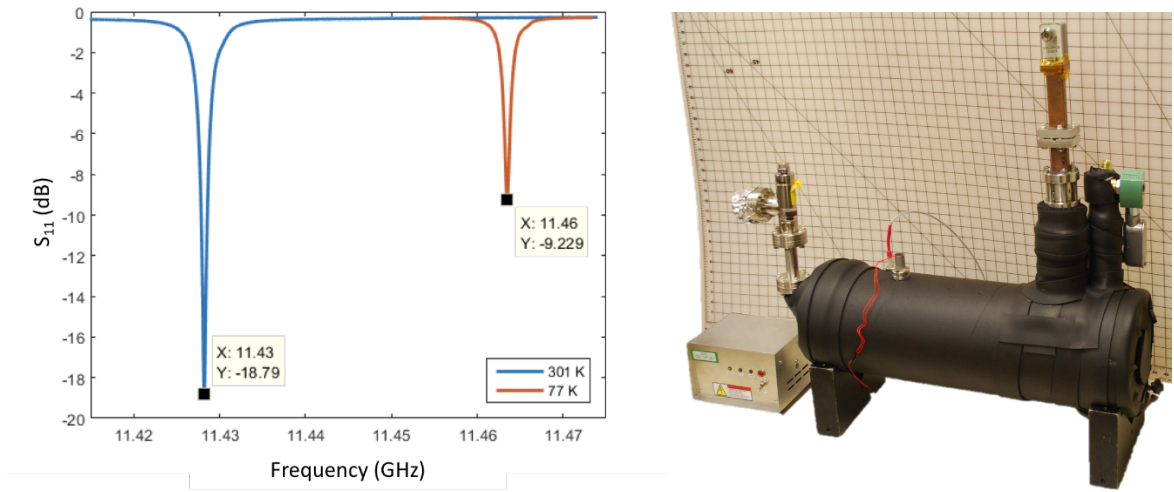

**Figure 16:** (left) S-parameters measured for an X-band 26 cm distributed coupling structure at both room temperature and 77 K, with the measured temperature provided in the legend. The structure transitioned from being under coupled to over coupled with a minimum $S_{11}$ of approximately -30 dB observed for critical coupling. (right) Photograph of the cryogenic vessel for an X-band distributed coupling structure. The background grid spacing is one inch.

The structure and the cryo-vessel have been installed at SLAC's NLCTA test facility for experiments to quantify the performance of the structure at high gradient. The structure is operated at a 60 Hz repetition rate, at a gradient of 120 MeV/m, and a flat-top pulse length of approximately 250 ns. The average thermal load in the cryo-vessel including thermal leakage for the foam-insulated stainless-steel cryogenic vessel and RF power deposited is estimated from cryogenic consumption to be 2 kW/m in proximity to the operational design for the main linac. The RF performance of the structure indicates that there is no degradation due to the cooling of the structure to remove the average power deposited. The expected temperature rise due to pulsed heating is 10 K on the cavity's inner surface. The structure can be operated stably at a gradient of 120 MeV/m with a constant flow of cryogens to replace losses due to evaporative cooling. The measured forward/reflected power, measured dark current, modeled reflected power and modeled gradient for the structure are shown in Figure 17.

## 7. Summary

We have presented a new concept for high gradient, high power linacs designed for an $e^+e^-$ linear collider in the TeV class. The linac design is based on two features: an accelerator structure with a separate feed to each cavity permitting the iris to be optimized for gradient and breakdown; and a structure that operates in LN, causing the Cu (or Cu alloy) conductivity to increase and reduce the RF power requirements by about a factor of 2.5. An initial analysis of both short-range and long-range wakefields for the selected parameters is presented. We observe that the short-range wakefield is tolerable with BNS damping and a small average off-crest operation. The long-range wakefield for the basic accelerating structure was found to be slightly less than the NLC design. This indicates that Gaussian-detuning and higher-order-mode damping manifolds may address this challenge. However, the detailed solution requires further design of the structure and the RF distribution. The optimal gradient for a cost-optimized collider depends strongly on the cost of RF power, and we assume the value from the DOE GARD decadal roadmap of $2/peak kW. This RF cost has not yet been demonstrated, but progress is being made on both klystrons and modulators, as will be reported elsewhere. If the DOE-HEP GARD[32] goal for RF power of $2/peak kW could be achieved, the linac cost and power, including tunnels and utilities but not $e^+e^-$ sources, would be about $3.2M/GeV and 171 kW/GeV, which is significantly less per GeV than other designs.

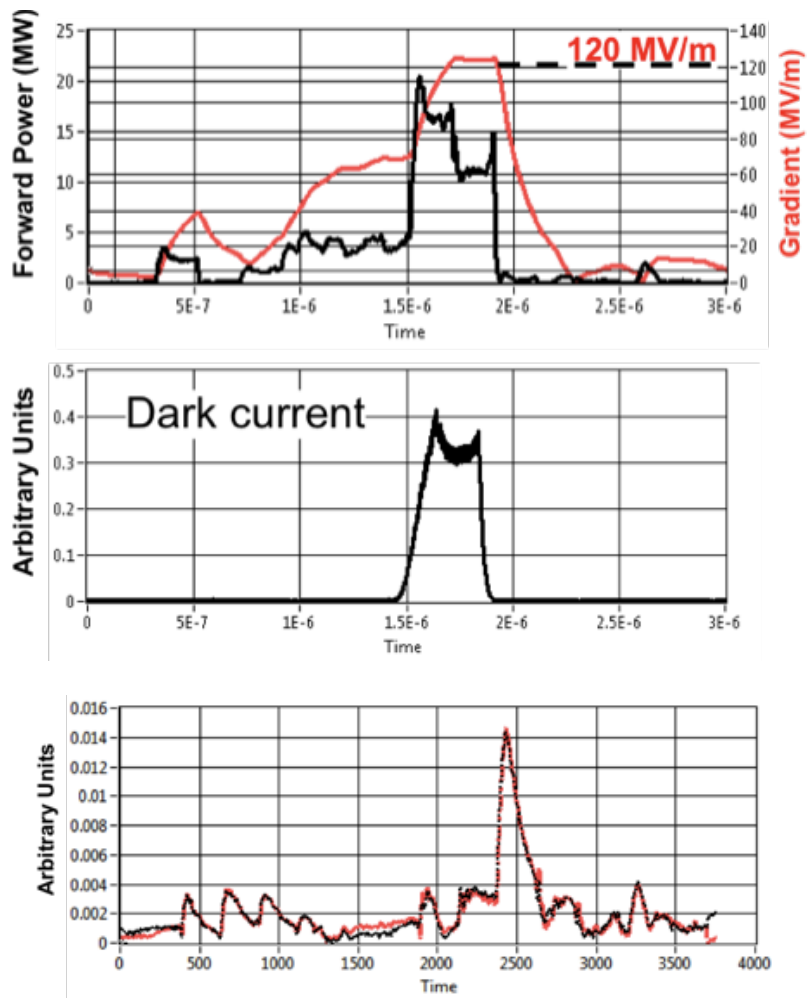

**Figure 17:** (Top) The measured forward power (black line) and modeled gradient (red line) for the structure operating at 120 MeV/m. The corresponding (middle) measured dark current, and (bottom) measured (black line) / modeled (red line) reflected power.

## 8. Acknowledgements

This work was supported by the Department of Energy under contract DE-AC02- 76SF00515 and the U.S.-Japan Science and Technology Cooperation Program under contract DE-AC02-76SF00515.

### Appendix A - Physics Case

The results from the LHC give renewed motivation to study $e^+e^-$ annihilation at high energy. First, the LHC experiments have discovered the Higgs boson and measured its mass. This discovery opens the door to the study of this last crucial element of the Standard Model of Particle Physics (SM) and defines the parameters for high-precision measurements of the properties of this particle. Second, the LHC has carried out extensive searches for new particles associated with models of fundamental physics. Thus far, no new particles have been found. Though there is still space to cover before reaching the ultimate limits of the LHC, the window is closing. However, the LHC cannot rule out particles with only electroweak interactions whose signatures at the LHC are subtle and elusive. There are important examples of such particles that are more easily discovered at $e^+e^-$ colliders.

These considerations lead to an important program of experiments to be carried out at future $e^+e^-$ colliders. To provide context, we briefly summarize these experiments in this section. They are now being studied for the physics programs of the ILC and CLIC. It is by no means assured that either of these colliders will go forward, or that whatever next $e^+e^-$ collider is built will reach multi-TeV energies. Still, the full set of experiments listed here should be carried out, and the technology described here may help.

A more complete description of these experiments, their motivation, and projections for achievable sensitivities can be found in the ILC Technical Design Report, in the ILC physics updates [5] and [33], and in the CLIC Conceptual Design Report.[34] The projections for precision Higgs boson measurements have been updated recently in References [35] and [36].

### a. Precision Measurement of Couplings Associated with Higgs Boson Decays

These are the Higgs boson couplings to $W$, $Z$, $b$, $\tau$, $c$, $\mu$, $\gamma$. It is now understood that measurements at 250 GeV center of mass with 2 ab$^{-1}$ and beam polarization (or 5 ab$^{-1}$ without beam polarization) can measure the first three of these couplings in a model-independent way to 1% accuracy or better. The search for deviations from the SM predictions in these couplings gives a new window into physics beyond the SM that is orthogonal to that of particle searches at the LHC. This is a must-do for particle physics. It is important to find an affordable technology and begin this program as soon as possible. Any anomalies discovered in this program could be confirmed by running at 500 GeV center of mass or higher, which would add measurements of $W$ fusion production of the Higgs boson. At 500 GeV, the individual coupling errors would be cut in half, and even better results could be expected at higher CoM's, using the cross section for $W$ fusion production of the Higgs boson which rises with increasing center of mass.

### b. Search for Exotic Decays of the Higgs Boson

The Higgs boson may couple to new particles with no SM gauge interactions, including the particle that makes up dark matter. Measurements at 250 GeV center of mass allow comprehensive searches for possible exotic modes of Higgs decay.[37]

### c. Precision Measurement of the Remaining Couplings of the Higgs Boson

Two important Higgs boson couplings require higher energy experiments. These are the Higgs coupling to the top quark and the Higgs self-coupling. In both cases, measurements of limited precision can be made at 500 GeV CoM, while higher-precision measurements require even higher energies. At 1 TeV in the center of mass, with 5 ab$^{-1}$ of data, it is possible to measure the $htt$ coupling to 2% accuracy and the $hhh$ coupling to 10%.[38] The latter analysis uses the complementary processes $e^+e^- \to Zhh$ and $e^+e^- \to \nu\nu hh$. It is argued in [39], that, although it is possible to measure double Higgs production in $pp$ collisions, it is very difficult to attribute a deviation from the SM prediction to a shift in the $hhh$ coupling. In fact, it is likely that, for both of these couplings, a precise, model-independent determination is possible only with an e$^+$e$^-$ collider.

### d. Precision Measurement of the Top Quark Mass

The top quark mass is a fundamental parameter of the SM, and so it is important to measure it as accurately as possible. For most applications, what is needed is a short-distance mass parameter, for example, the $\overline{MS}$ mass. At this moment, it is not understood how to convert the top quark mass quoted by the LHC experiments to a short-distance mass, or even to the top quark pole mass. The conversion from the pole mass to the $\overline{MS}$ mass brings in a perturbative theoretical uncertainty of 200 MeV,[40] plus less well characterized nonperturbative uncertainties. On the other hand, the position of the $t\bar{t}$ threshold in e$^+$e$^-$ annihilation is controlled by a mass parameter that is very close to the $\overline{MS}$ mass. Measurement of this threshold would give the top quark mass to an accuracy of 40 MeV.[5]

### e. Precision Measurement of the Top Quark Electroweak Couplings

Models in which the Higgs boson is composite typically predict sizable deviations from the SM expectations for the $W$ and $Z$ couplings of the top quark. At an e$^+$e$^-$ collider, the $Z$ vertices appear in the production reaction, where the individual contributions of $s$-channel $\gamma$ and $Z$ exchange can be disentangled using beam polarization. This allows measurements of the $Z$ vertices with precisions of better than 1%.[5] These effects are proportional to $s/m_Z^2$ and so are increasingly visible at higher energies.

### f.  Search for Pair-Production of Invisible Particles

Though much attention is being given today to searches for pair-production of invisible particles (including dark matter particles) at the LHC, this is a difficult endeavor. If the dark particles are connected to SM particles by a hard operator or a heavy-mass exchange, it is possible to search for very high $pT$ monojet events. However, if the production is by electroweak Drell-Yan production, the mass reach is limited by QCD and parton distribution uncertainties in the estimation of the irreducible background process, Drell-Yan production of $Z \rightarrow \bar{\nu}\nu$. For example, for Higgsinos, the discovery reach of LHC is expected to be less than 200 GeV.[41] In contrast, the pure Higgsino is thermally produced as dark matter with the correct relic density at a mass of about 1 TeV.[42]

In e$^+$e$^-$ annihilation, the corresponding process of photon plus missing momentum is much more precisely understood, allowing searches for invisible particles almost to the kinematic limit.[43] Thus, an e$^+$e$^-$ collider operating above 2 TeV center of mass might be the unique way to search for this special dark matter candidate and other high-mass invisible particles.

### g.  Search for New Electroweak Gauge Bosons and Lepton Compositeness

The processes e$^+$e$^-$ $\rightarrow f\bar{f}$ can be used to search for new electroweak gauge bosons and for 4-fermion contact interactions including those signaling fermion compositeness. For 3 TeV center of mass and 1 ab$^{-1}$ of data, an e$^+$e$^-$ collider would be sensitive to $Z'$ masses of 15 TeV, more than double the eventual LHC reach. Such a collider would be sensitive to compositeness scales in the range of 60-80 TeV.[34]

Thus, there are strong physics arguments to construct a new, high-energy e$^+$e$^-$ collider spanning the energy range from 250–350 GeV to multi-TeV CoM.

### Appendix B - Main Linac Optimization

A model was developed to estimate the cost, $C,$ of the main linac for a linear collider given by $C \approx xE/G + yE(I + G/R_s)$ where $E$ is the center of mass beam energy; $x$ is the accelerator cost/length including the tunnel outfitting, cryomodule and structure costs; $G$ is the accelerating gradient; $y$ is the RF system cost/peak power including the power supplies, modulator, high power amplifier, etc.; $R_s$ is the shunt impedance of the structure; and $I$ is the beam current. These quantities were not fixed, but rather they depended on the mode of operation for the accelerator. For example, the RF source cost was increased from the nominal $/kW value as the duty factor of the accelerator increased to account for additional power handling, thermal cooling and energy

storage in the RF source. Cost estimates for building structures and outfitting tunnels were varied based on the requirements for different temperatures of operation to account for the inclusion of cryogenic systems and cryoplants. The structure cost covers: fabrication of the accelerator structure; quadrupole magnets; supports; vacuum; the active alignment and magnet stabilization system; and the cryostat and cryogenic distribution if required by the temperature of operation. The tunnel costs include boring, water distribution and outfitting with electrical power. As the structure's aperture and the electron bunch charge are similar to CLIC, we anticipate the need for transverse tolerances and an active stabilization system similar to the CLIC design.[44] Finally, specific structures were designed and modeled at C-band and X-band to directly compare practical shunt impedances.

In Figure 18 (left) the cost of the main linac is shown as a function of beam loading and structure shunt impedance for an operating temperature of 77 K. This operating temperature includes additional cost for cryoplants and cryogenic housing for the accelerator as described in Table 4 and Table 2, respectively. This figure illustrates how both extremely low and extremely high beam loading drive up the cost. For very high beam loading, the peak power requirements of the RF sources drive up the cost; for low beam loading, the average power requirements of the RF sources become the dominant factor. As a second constraint we elected to operate with a site power of no greater than 250 MW for the main linacs in order to not exceed practical overall requirements for the accelerator complex. This power requirement sets a minimum beam loading of 42.5% that corresponds to the upper limit of the region where cost scales slowly with beam loading. The optimal gradient for each operating point of Figure 18 (left) is shown in Figure 18 (right) where we see that lower beam loading and higher structure impedance allows for a cost optimized operation at higher gradient.

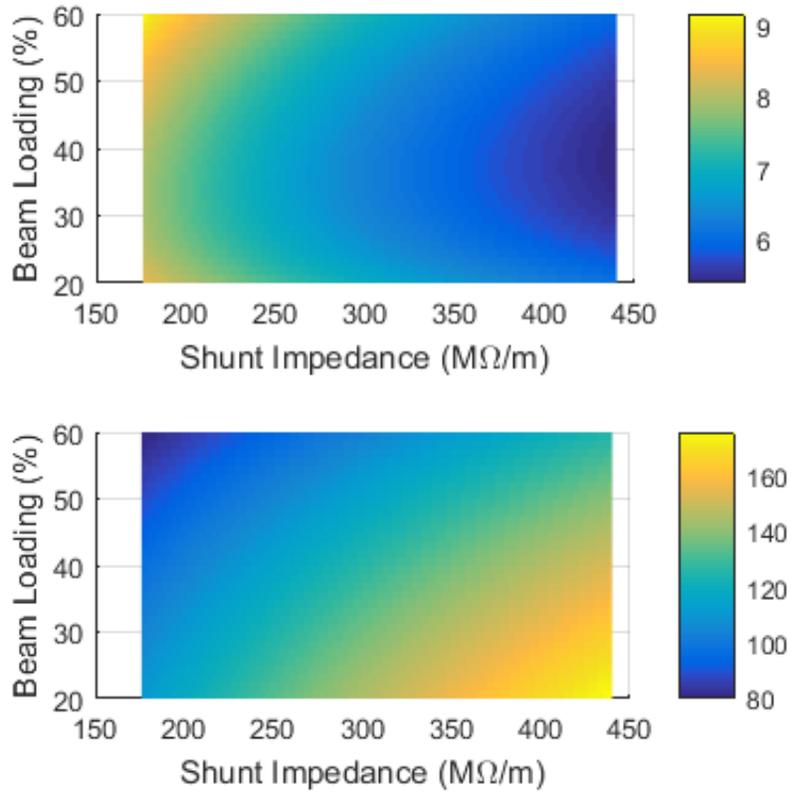

**Figure 18:** (top) Cost [G$] and (bottom) accelerating gradient [MeV/m] of two 1 TeV linacs operating at 77 K as a function of beam loading and structure shunt impedance.

The RF source cost (including the RF amplifier and modulator) is a function of the pulse length and duty factor required to achieve the prescribed beam loading. This cost was varied from the nominal $2/kW-peak due to power handling constraints using an empirical scaling derived from commercial units, as shown in Figure 19. The empirical scaling was a function of $y=y_0(A_1\tau^2+A_2\tau+A_3)$ where $y_0$ is the nominal cost, $\tau$ is the pulse length, $A_1$ = 0.96838, $A_2$ = 0.3162, $A_3$ = 2.9399e-05.

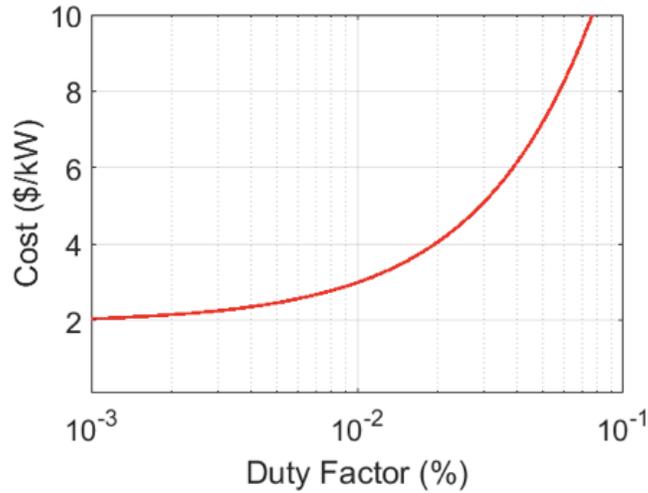

**Figure 19:** Cost scaling of RF sources as a function of duty factor derived from empirical formula.

For the selected operational point described in Table 2 we show in Figure 20 the annual electrical power cost of operation and main linac power requirements. The strong dependence observed due to beam loading renders designs with low beam loading cost prohibitive and impractical to implement with existing electrical power infrastructure. Therefore, a limit of 350 MW for the main linac was selected for this design.

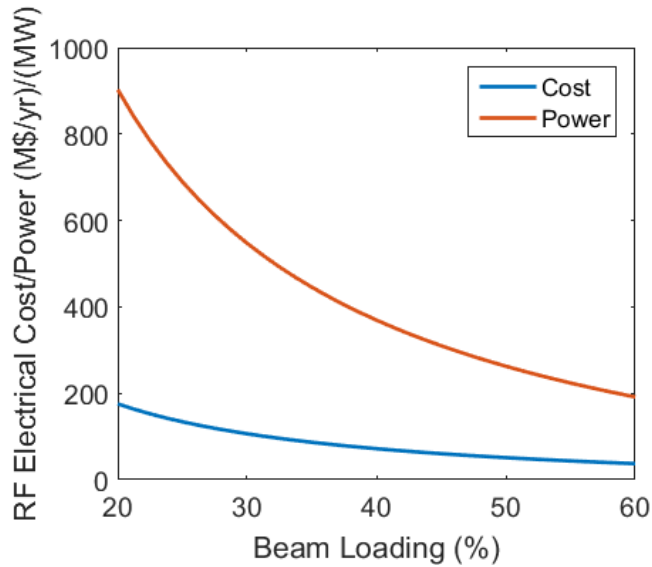

**Figure 20:** The annual electrical operating cost (blue) and power requirements (red) for the main linac, based on $10^7$ s/yr of operation a 7 cents/kw-hr electrical power contract.

The optimization of the system for various RF source costs, $y_0$, results in different operational regimes for the collider. Table 5 shows how this variation impacts the cost and performance on the collider for various scenarios constrained by a site electrical power of 350 MW.

| RF Source Cost ($/kW) | 2 | 4 | 6 | 12 |
|---|---|---|---|---|
| Effective RF Source Cost Including Duty Scaling ($/kW) | 2.2 | 4.4 | 6.6 | 13.3 |
| Temperature | 77 | | | |
| Main Linac Cost (G$ for 2 TeV COM) | 6.4 | 8.9 | 10.8 | 14.9 |
| Structure Cost (k$/m) | 100 | | | |
| Beam Loading (%) | 42.5 | | | |
| Gradient (MeV/m) | 117 | 87 | 71 | 51 |
| Flat Top Pulse Length (µs) | 0.25 | 0.35 | 0.43 | 0.61 |
| Electrical Load from Cryogenics (MW) | 170 | | | |
| Electrical Load (MW) | 342 | | | |

**Table 5:** Optimized operating points as a function of RF source cost.

**Appendix C – Beam Tunnel Aperture**

The beam tunnel aperture of the accelerating structure plays a significant role in determining the structure shunt impedance, with smaller apertures producing higher shunt impedances. However, a larger iris is preferred as the iris aperture plays a strong role in determining the strength of the electron bunch's wakefields. Multiple structures at C-band and X-band were designed and simulated in order to make an initial estimate of the tolerable aperture for the structure. The operating parameters, *e.g.* gradient, bunch charge, bunch length, etc., for these designs would not be identical, but would require a full investigation of each design to determine. Therefore, the comparison between structures was performed for a fixed set of parameters. In Figure 21, the residual energy spread is shown as function of electron bunch charge for a fixed operating point of 100 MeV/m, a bunch length $\sigma_z = 150\ \mu m$, and an offset in phase of 25 degrees. In this plot the increase in residual energy spread a low charge corresponds to operating with a phase offset that is too large and should be reduced. It is clear that the performance of all structures with an aperture of 2.63 mm or larger allows for a residual energy spread of significantly less than 1%. Operation with a smaller aperture would not be possible for <1% energy spread and a 1 nC bunch charge.

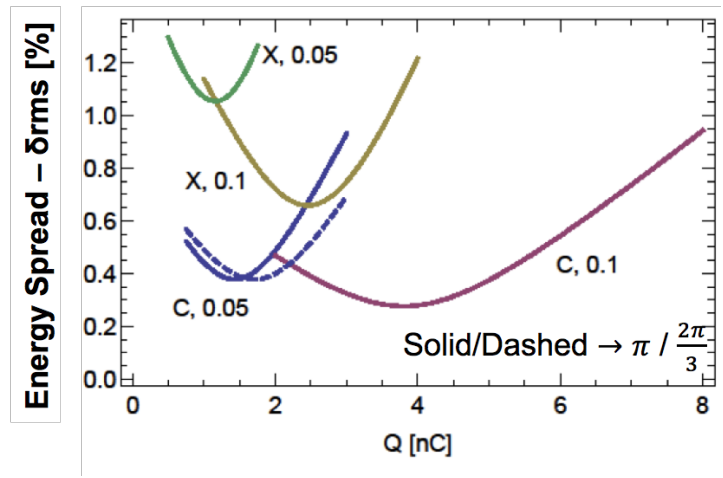

**Figure 21:** Residual energy spread is shown as function of electron bunch charge for a fixed operating point of 100 MeV/m, a bunch length $\sigma_z$ = 150 $\mu$m, and an offset in phase of 25 degrees. Solid/dashed lines correspond to $\pi$ / $2\pi$/3 mode structures. The structure's frequency and aperture are listed adjacent to each line and correspond to the shunt impedance listed in Table 6.

| Frequency | $a/\lambda$ | Phase Advance | $R_s$ (M$\Omega$/m) |
|---|---|---|---|
| C-band (5.712 GHz) | 0.05 | $\pi$ | 121 |
| C-band (5.712 GHz) | 0.05 | $2\pi$/3 | 133 |
| C-band (5.712 GHz) | 0.1 | $\pi$ | 92 |
| X-band (11.424 GHz) | 0.05 | $\pi$ | 176 |
| X-band (11.424 GHz) | 0.1 | $\pi$ | 133 |

**Table 6:** Optimized structure shunt impedance at 300 K.

**Appendix D – Comparison of RF Pulse Formats**

Due to the high Q-factor of the cryogenic cavities, the resulting fill time requires for a significant amount of additional RF power. It is possible however that the filling of the cavities can have a minor impact on the thermal loading for the cryogenic cooling systems with appropriate tailoring of the pre-pulse. Recent advances in pulse compressors[45,46,47] with super-compact spherical cavities has dramatically changed the performance potential for these systems. A chain of spherical cavities can produce tailored pulse formats with extremely high conversion efficiency if the power gain is kept low.[48] A pre-pulse power compression ratio of three can be achieved with >80% conversion efficiency. For the baseline mode of operation, we have explored the impact of operating with and without a compressed pre-pulse on the thermal and electrical power requirements for the linac.

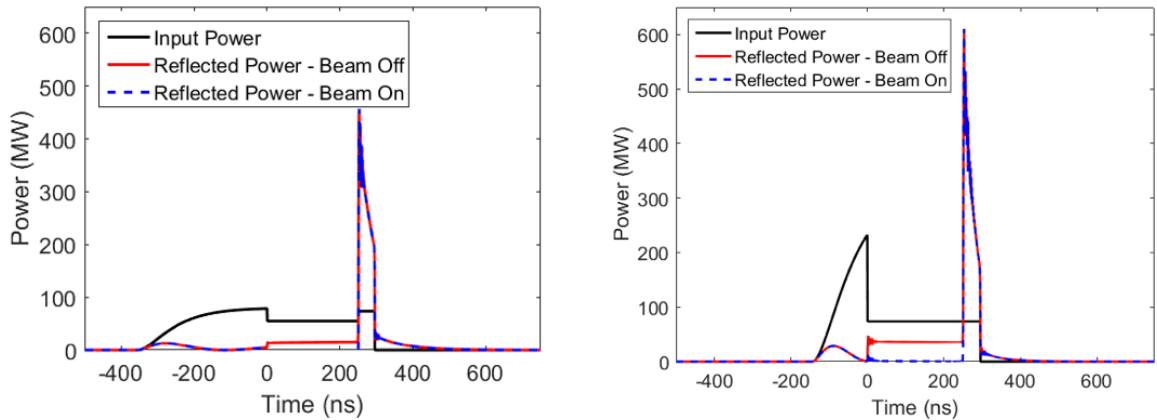

**Figure 22:** The forward and reflected RF power for a (left) flat input pulse that does not exceed 80 MW and a (right) 3X power compressed pre-pulse. The RF traces in black and red assume no beam loading. The blue trace is the reflected power with beam loading and 80 MW delivered during the flat top of the RF pulse.

In Figure 22 we compare the RF power traces for two modes of operation. One where the maximum power out of the klystron matches the power into the linac referred to as the "flat pulse". The second with a pre-pulse that reaches nearly three times the forward power of the klystron, referred to as "compressed pre-pulse". A short phase-inverted post pulse is utilized in both cases to dump the fields in the cavity. This post-pulse saves roughly 10 MW of overall electrical power while reducing the thermal load in the cryogenic system. Both systems are optimized with highly over coupled cavities due to the heavy beam loading. The resulting gradient is shown in Figure 23 with the "flat pulse" on the left and the "compressed pre-pulse" on the right.

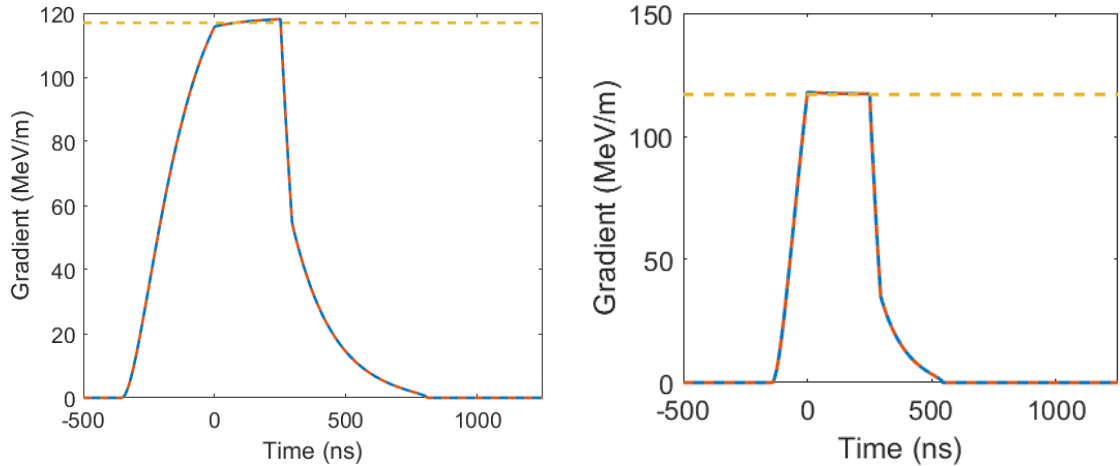

**Figure 23:** The gradient in the cavity during a (left) flat input pulse and (right) three times power compressed pre-pulse.

In Table 7 a comparison is made between the power requirements for the "flat pulse" and the "compressed pre-pulse". Significant electrical power savings are possible with the inclusion of a pulse compressor. This is due to the reduced power loss in the cavity during the long fill time of the cavity, which reduces both the klystron electrical power requirements, as well as the cryogenic cooling requirements. The difference in capital costs between the two systems as the cryogenic cooling costs are replaced by the cost of compressors, however the reduced electrical power makes the latter significantly more favorable.

In addition, we have also considered operating with a compressed pulse during the flat-top portion of the RF pulse. This does not prove to be realistic for the power ratios required during the pre-pulse of the system to maintain reasonable power consumption. Increasing the compression for the full pulse by an additional factor of 2 (6X for the pre-pulse) reduces the overall efficiency of the system by a factor of three. Roughly this would require an additional 300 MW of electrical power for the system doubling the overall electrical consumption of the collider. This is a startling result considering that pre-pulse compression actually increases the system efficiency by 120 MW for overall power consumption. In practice this is due to the rather modest requirements of pre-pulse compression becoming excessive in the presence of flat-top compression.

| Parameter (2 TeV CoM) | Units | Value | |
|---|---|---|---|
| Nitrogen Reliquification Plant Cost for Cryogenic Cooling | M$/MW | 18 | |
| Pulse Shape | | Flat-Pulse | Compressed Pre-Pulse |
| Single Beam Power (1 TeV linac) | MW | 9 | |
| Total Beam Power | MW | 18 | |
| Total RF Power | MW | 105 | 86 |
| Heat Load at Cryogenic Temperature | MW | 38 | 25 |
| Electrical Power for RF | MW | 210 | 172 |
| Electrical Power for Cryo-Cooler | MW | 254 | 170 |
| Total RF Power | MW | 464 | 342 |

**Table 7:** Power consumption breakdown by major linac system.

**References**

_______________________